\documentclass[twocolumn,showpacs,preprintnumbers,amsmath,amssymb,aps]{revtex4}
\usepackage{epsf}
\usepackage{graphicx}
\usepackage{dcolumn}
\usepackage{bm}

\begin{document}


\title{Quantum evolution of the Universe from $\tau=0$.}

\author{S.L. Cherkas}
\affiliation{Institute of Nuclear Problems, 220050 Minsk, Belarus}
\date{\today}

\begin {abstract}
We have shown that DeWitt constraint $H=0$ on the physical states
of the Universe does not prevent Heisenberg operators and its mean
values to evolve with time.  Mean value from observable, which is
singular in classical theory, is also singular in a quantum case.
\end{abstract}
\maketitle

\section{Introduction}
One of the first attempts to quantizy gravity were made by DeWitt
\cite{witt}. Then his equation $H\psi=0 $, where $H$ is
Hamiltonian, was applied for quantizing cosmology. In the simplest
case of isotropic and uniform Universe filled with scalar field
the equation contains two variables: scale factor of the Universe
and amplitude of the scalar field. There is no "time" in this
equation, whereas we are interested namely in the Universe
evolution in time. This leads to various discussions about "time
disappearance" and interpretation of wave function of Universe
\cite{hall}. Possible solutions like to introduce time along
quasiclassical trajectories, or subdivide Universe into classical
and quantum parts were submitted \cite{vil}. Such point of view
can not be satisfactory. Ideally, time must exist independently of
we consider Universe quantum or classically. Let us remind that
this situation is analogous to that in string theory, where
constraint H=0  also exists. Nevertheless this constraint does not
prevent evolution of Heisenberg operators $\hat X(\sigma,\tau)$
from $\tau$.

Very short we can describe the problem as following. Let we have
wave function $\psi(a,x)$ depending on two variables which
satisfies
\begin{equation}
\hat H\psi(a,x)=0.
\end{equation}
For evolution of some Heisenberg operator $\hat H$  we have

\begin{equation}
<A(\tau)>=<\psi|e^{i\hat H\tau}\hat A e^{-i\hat H\tau} |\psi>.
\end{equation}
At first sight one sees no evolution, but it is not true for this
case. Really $e^{-i\hat H\tau} |\psi>=|\psi>$, but we can't write
$<\psi|e^{i\hat H\tau}=<\psi|$. The point is that the wave
function can not be normalized the by an ordinary way $\int
\psi^*(a,x)\psi(a,x)dxda=1$ due to constraint. In fact the
function is unrestricted along one of the variable. For instance,
it is $a$ variable. So if $H$ contains differential operator like
$\frac{\partial^2}{\partial a^2 }$ we can not remove
$\frac{\partial^2}{\partial a^2}$ to the left by habitual
operation $<\psi|\frac{\partial^2}{\partial a^2
}=<\frac{\partial^2}{\partial a^2 }\psi|$ through the integration
by parts. As a result, is not true to
 apply that
$<\psi|\hat H=<\hat H\psi|=0$ is valid.

 Thus we need to
solve equations for Heisenberg operators in quantum cosmology to
obtain evolution of the Universe. In general case, solving of the
equations for Heisenberg operators is hopeless problem, still, for
the simplest case of two variables it be done.

\section{Relativistic particle}

The problem of quatazation of cosmology have many similar features with that
 of the relativistic particle \cite{mori}.
 The action of the relativistic particle has the form:

\begin{equation}
S=-m\int\sqrt{-\dot{x}_\mu^2}d\tau \label{s1}.
\end{equation}
The equivalent form, from which (\ref{s1}) can be obtained by the
varying on the lapse function $e(\tau)$
 is the

\begin{equation}
S=\frac{1}{2}\int(e^{-1}\dot{x}_\mu^2-e \,m^2)d\tau. \label{s2}
\end{equation}
One more equivalent form, from which(\ref{s2}) arises after
varying $p_\mu$, has the form:
\begin{equation}
S=\int\bigl\{p_\mu
\dot{x}^\mu-e\left(\frac{p_\mu^2+m^2}{2}\right)\bigr\}d\tau.
\end{equation}

Using the reparametrisation invariance \cite{kaku} we can choose
the lapse function equals to $e=1/m$, so from the last equation
one can see that Hamiltonian is equal to
$H=\frac{p_\mu^2+m^2}{2m}$ and, besides, varying on $e$ gives
constrain  $H=0$. After quantization
\[
[\hat p^\mu,\hat x^\nu]=ig^{\mu\nu}, ~~~~\hat p^\mu=\{\hat
\varepsilon, \hat{ \bm p} \}\equiv\{i\frac{\partial}{\partial
t},-i\bm \nabla\}
\]
the constrain becomes the Klein-Gordon equation. Commuting the
position  and four-momentum operators with the Hamiltonian one
obtains Heisenberg equation of motion:
\begin{eqnarray*} \frac{d \hat
x^\mu}{d\tau}=i[\hat H,\hat x^\mu]=\frac{\hat
p^\mu}{m}, \nonumber\\
\frac{d\hat  p^\mu}{d \tau}=0
\end{eqnarray*}

Evident solution of the equation of motion is
\begin{eqnarray*}
\hat x^\mu(\tau)=x^\mu+\frac{p^\mu(0)}{m}\,\tau\\ \hat
p^\mu(\tau)=\hat p^\mu,
\end{eqnarray*}
where $\hat x(0)=x$ and $\hat p(0)=\hat p$. As we see two "times"
appears in the above equations. We shall refer to the $\hat
t(\tau)$ as "physical time" and to the $\tau$ as a "proper time".
Physical time is operator, whereas proper time is some parameter
"always running forward" like time in Newtonian physics.

One can easy check that the wave function satisfying Klein-Gordon
equation can't be normalized through the integration $d^4x$
\cite{vil}, however, it can be normalized through the time-like
component of the conserved four-current:
\[
<\psi|\psi>=-i\int
\{\psi(\bm r,t)\partial_t\psi^*(\bm r,t)-(\partial_t\psi(\bm r,t))\psi^*(\bm r, t)\}d^3\bm r.
\]

Normalized wave packets have the form
\[
\psi(x)=\int \frac {a(\bm k)e^{-i\varepsilon(\bm k)t+i\bm k\bm r}}
{\sqrt{2\varepsilon(\bm k)(2\pi)^3}}d^3\bm k;~~~\int \mid a(\bm
k)\mid^2d^3\bm k=1,
\]
where $\varepsilon(\bm k)=\bm k^2+m^2$. To avoid appearance of the
states with negative norm we must take only positive frequency
solutions.
 Using this wave
packet we see that integration $\int \psi^*(x)\psi(x)d^4x$ does
not converge and conclude that the wave function is not restricted
over the $t$ variable. As a consequence,  Heisenberg operators
evolve with $\tau$ despite of constrain $H\psi=0$.

Define now mean value of some operator $A$ as

\begin{eqnarray}
<\hat A(\tau)>=i\int \biggl(\frac{\partial \psi}{\partial
t}(\hat A(\tau)\psi)^*~~~~~~~~\nonumber\\
-\left(\frac{\partial \psi^*}{\partial t }\right)\hat
A(\tau)\psi\biggr)d^3\bm r \bigr| _{t=0}. \label{mean}
\end{eqnarray}
Let us note, that after integration over $d^3\bm r$ in
(\ref{mean}) we should set $t=0$.
 This definition has the following properties:
\par \noindent
 1) It is consistent
with the normalization of wave function if we choose $\hat A$ to
be equal to the unit operator.
\par \noindent
2) It look like as an expression for the Heisenberg operator mean
value in the nonrelativistic picture when the operator acts on the
wave function taken at the initial moment of time.
\par\noindent
3)It has natural  property $\frac{<\hat
A(\tau)>}{d\tau}=<\frac{\hat A(\tau)}{d\tau}>$
\par\noindent
4)It gives physical time value, equal to zero, when proper time
equals to zero.

Averaging of the Heisenberg equation of motion gives
\begin{eqnarray*}
<\hat t(\tau)>=<\hat \varepsilon> \frac{\tau}{m},\\
\hat {\bm r}(\tau)=<\hat{\bm r}>+ <{\bm p}>\frac{\tau}{m}.
\end{eqnarray*}
We see that physical time goes proportional to the proper time. It
is interesting to calculate the dispersion of the physical time
for the Gaussian wave packet $a(\bm k)\sim e^{-\bm k^2}$ with the
$\frac{1}{\alpha}>>m^2$. Evaluating of the mean values of the
$\hat \varepsilon =i\frac{\partial }{\partial t}$ and its square
gives:
\begin{eqnarray*}
 <\hat
\varepsilon>=\frac{\int_0^\infty \varepsilon (k)e^{-\alpha k^2
}k^2dk}{\int_0^\infty e^{-\alpha k^2
}k^2dk}\approx\frac{2}{\sqrt{\pi\alpha}}\\
<\hat \varepsilon^2>=\frac{\int_0^\infty \varepsilon^2
(k)e^{-\alpha k^2 }k^2dk}{\int_0^\infty e^{-\alpha k^2
}k^2dk}\approx\frac{3}{2\alpha},
\end{eqnarray*}
and we come to
\begin{eqnarray*}
\frac{\sqrt{<\hat t^2>-<\hat t>^2}}{<\hat t>}=\frac{\sqrt{<\hat
\varepsilon^2>-<\hat \varepsilon>^2}}{<\hat
\varepsilon>}=\sqrt{\frac{3\pi}{8}-1}.
\end{eqnarray*}
Thus, "particle-clock" is a bad clock when particle localized in
the region less than Compton wave length. Let us imagine what is
to be when we place this particle (for instance having electric
charge) in the electric field. For this aim we consider classical
picture do not having straight relation to the quantum picture
but, still correctly it reflecting. Let we have cloud of the
particles with some initial dispersion of the energy placed in the
electric field. The particles begin to accelerate in the electric
field and receive the energies much larger than its initial
energies, so that the energy dispersion becomes negligible.
Similar picture holds for the quantum case so that the
"particle-clock" from "bad clock" becomes "Swiss clock".

\section{Quantum cosmology}
We start from the Einstein action of gravity and the action of a
one-component real scalar field:
\begin{equation}
S=\frac{1}{16\pi G}\int d^4 x\sqrt{-g}R+\int d^4
x\sqrt-g[\frac{1}{2}(\partial_\mu\phi)^2-V(\phi)],
\end{equation}
where $R$ is a scalar curvature and $V$ is a matter potential
which includes a possible cosmological constant effectively. We
restrict our consideration to the homogeneous and isotropic
metric:
\begin{equation}
ds^2=N^2(\tau)d\tau^2-a^2(\tau)d\sigma^2.
\end{equation}
 Here the lapse function
 $N$ represents the general time coordinate transformation
 freedom.
For the restricted metric the total action becomes
\begin{eqnarray*} S=\int N(\tau)\biggl\{ \frac{3}{8\pi G}a\left(
{\mathcal K}-\frac{\dot{a}^2}{N^2(\tau)} \right)~~~~~~~~~
\\+\frac{1}{2}a^3\frac{\dot{\phi}^2}{N^2(\tau)}-
a^3V(\phi)\biggr\}d\tau,
\end{eqnarray*}
where $\mathcal{K}$ is the signature of the spatial curvature.
 This action can be obtained from the following
expression by varying over $p_a$ and $p_\phi$
\begin{eqnarray*}
S=\int\biggl\{p_\phi\dot{\phi}+p_a\dot{a}+N(\tau)
\biggl(-\frac{3a{\mathcal K}}{8\pi G}-\frac{8\pi G\,p_a^2}{12a}
\\
+\frac{p_\phi^2}{2a^3}+a^3V(\phi)\biggr)
    \biggr\}d\tau.
\end{eqnarray*}
Varying on $N$ gives the constraint
\[
H=-\frac{3a{\mathcal K}}{8\pi G}-\frac{8\pi
G\,p_a^2}{12a}+\frac{p_\phi^2}{2a^3}+a^3V(\phi)=0.
\]
After quantization $[\hat a,\hat p_a]=i$, $[\hat \phi,\hat
p_\phi]=i$ this constraint turns into the DeWitt equation
\[
\hat H\psi(a,\phi)=0.
\]
Every time someone has scratched hands to modernize DeWitt
constrain (as, for instance, in \cite{lasuk}). This immediately
implies (as in classical so in quantum cases) existence of some
preferred system of reference. Although there are some logically
consistent theories implying preferred system of reference, for
instance, Logunov relativistic theory of gravity \cite{log},
giving adequate description of all the stages of Universe
expansion \cite{kalash}, still, we prefer to keep to General
Relativity here and do not touch the constraint.

After quantization we should to define operator ordering in
Hamiltonian. For the first time we take the given operator:
\begin{equation}
\hat H=-\frac{1}{4}\left({\hat p_a}^2\frac{1}{a}+\frac{1}{a}{\hat
p_a}^2\right)+\frac{\hat p_\phi^2}{2 a^3}-\frac{{\mathcal
K}a}{2}+a^3V(\phi).
\end{equation}
In this equation we use system of units which reduce the number of
constants. Let us to consider the wave function in the vicinity
$a=0$. When $a\rightarrow 0$ the terms $a^3V(\phi)$ and ${\mathcal
K }a$ do not influence on the form of the wave function and it
satisfy to the $H_0\psi=0$, where the simplified Hamiltonian
equals
\begin{equation} \hat H_0=- \frac{1}{4}\left({\hat
p_a}^2\frac{1}{a}+\frac{1}{a}{\hat p_a}^2\right)+\frac{\hat
p_\phi^2}{2 a^3}.
\label{hamilt}
\end{equation}
Explicit expression for the wave function is
\[
\psi(a,\phi)=a^{1\pm i|k|}e^{ik\phi}.
\]

Exactly as in the case of the Klein-Gordon equation we should
choose only positive frequency solutions. Thus the  wave packet
\begin{equation}
 \psi(a,\phi)= \int c(k)\frac{a^{1-
i|k|}}{\sqrt{4\pi|k|}}e^{ik\phi}d k. \label{pack}
\end{equation}
will be normalized by
\begin{equation}
i\int\left(\left(\frac{1}{a}\frac{\partial \psi}{\partial
a}\right)\psi^*-\left(\frac{1}{a}\frac{\partial \psi}{\partial
a}\right)^* \psi \right)d\phi=\int c^*(k)c(k)dk=1. \label{nm}
\end{equation}
As we can see $a$ variable plays here role of the physical time,
whereas $\tau$ is proper time.
  After finding of the evolution of the operators in the
in the proper time
   $
\frac{d\hat A(\tau)}{d\tau}=i[\hat H,\hat A], $ we find mean it
value:
\begin{eqnarray}
<A(\tau)>=i\int \biggl(\frac{1}{a}\frac{\partial \psi}{\partial
a}(\hat A(\tau)\psi)^* ~~~~~~~~~~~~\nonumber\\
-\left(\frac{1}{a}\frac{\partial \psi}{\partial a }\right)^*\hat
A(\tau)\psi\biggr)d\phi\biggr|_{a=0}. \label{geiz}
\end{eqnarray}

To find evolution for small $\tau$ we again may to use simplified
Hamiltonian $\hat H_0$. Evaluation of the commutators gives
\begin{eqnarray}
{\hat p}_\phi^{\bm\cdot}(\tau)=0;~~~{\hat p}_\phi(\tau)=const;\label{eqa}\\
(\hat a^3(\tau))^{\bm\cdot}=-\frac{3}{2}(\hat p_a  a+ a \hat p_a);\label{eqb}\\
(\hat p_a  a+ a \hat p_a)^{\bm\cdot}=-6\hat H_0. \label{eq}
\end{eqnarray}
From the equation (\ref{eq}) one may conclude that the third
derivative from the $\hat a^3(\tau)$ is already equal to zero, so
$\hat a^3(\tau)=a^3+\hat D\tau+\hat B\tau^2$. Using (\ref{eqa}),
(\ref{eqb}), (\ref{eq}) allows to find operator constants $\hat D$
and $\hat B$:
\begin{eqnarray}
\hat a^3(\tau)=a^3-\frac{3}{2}(\hat p_a  a+ a \hat
p_a)\tau~~~~~~~~ \nonumber\\
+\left(\frac{9}{8}\left(\frac{1}{a}\hat p_a^2+\hat
p_a^2\frac{1}{a}\right)-\frac{9}{4}\frac{\hat p_\phi^2
}{a^3}\right)\tau^2 \label{6},
\end{eqnarray}
where $\hat p_a$, $\hat p_\phi$ are not Heisenberg operators, but
ordinary  $\hat p_a=-i\frac{\partial}{\partial a}$, $\hat
p_\phi=-i\frac{\partial}{\partial \phi}$.

Evaluation of the mean value over wave packet gives
\begin{equation}
<a^3(\tau)>=3\tau\int(\frac{3}{2|k|}+|k|)|c(k)|^2 dk.
\end{equation}
It similar to the classical evolution given by the equations:
\begin{eqnarray}
-\frac{p_a^2}{2a}+\frac{p_\phi^2}{2a^3}=0;\nonumber\\
a^\cdot=-\frac{p_a}{a};~~~\phi^\cdot=\frac{p_\phi}{a^3};\nonumber\\
p_a^\cdot=-\frac{p_a^2}{2a^2}+\frac{3}{2}\frac{p_\phi^2}{a^4};\nonumber\\
p_\phi=const.
\end{eqnarray}
Classical evolution gives $a^3(\tau)=3|p_\phi|\tau$.

Next interesting quantity is mean value from the  $\frac{\hat
p_\phi^2}{2 \hat a^3}$ which ruffly is energy density of scalar
field. In classics the quantity looks like
\begin{equation}
\frac{p_\phi^2}{2a^3(\tau)}=\frac{|p_\phi|}{6\tau}
\end{equation}
and goes to infinity when $\tau\rightarrow 0$.

For the quantum case we must to find the action of the operator
$\frac{1}{ \hat a^3(\tau)}$ to  the wave function i.e. to solve
the equation $\hat a^3(\tau)\Theta(\tau,a,k)=a^{1-i|k|}$, and find
$\Theta(\tau,a,k)=\frac{1}{ \hat a^3(\tau)}a^{1-i|k|}$.

Explicit solution of the equation looks like
\begin{widetext}
\begin{equation}
\Theta(\tau,a,k)=\frac{a^{1-i|k|}}{3|k|\tau}-\frac{a^{1-i|k|}}{3|k|\tau}
\left(\frac{2ia^3}{9\tau}\right)^{\frac{2i}{3}|k|}e^{\frac{2ia^3}{9\tau}}
       \Gamma(1 - \frac{2\,i }{3}\,|k|,\frac{2\,i
       \,a^3}{9\tau })
\label{thet}
\end{equation}
\end{widetext}
where $\Gamma(a,z)=\int_z^\infty t^{a-1}e^{-t}dt$ is incomplete
Gamma function. Finding of the mean value of the $<\frac{\hat
p_\phi^2}{\hat a^3(\tau)}>$ reduces to the evaluation of the
\begin{widetext}
\begin{equation} <\frac{\hat p_\phi^2}{\hat a^3(\tau)}>=i\int
((1+i|k|)a^{-1 + i \,|k|}\Theta(\tau,a,k)-(1-i|k|)a^{-1 - i
\,|k|}\Theta^*(\tau,a,k))|k||c(k)|^2dk\biggr|_{a\rightarrow 0}.
\end{equation}
\end{widetext}
Analysis of the asymptotic of the (\ref{thet}) shows that second
item containing the Gamma function gives zero contribution after
the integration over $k$ and proceeding to the limit $a\rightarrow
0$ for any normalizable $c(k)$. As a result we have
\begin{equation}
<\frac{\hat p_\phi^2}{\hat a^3(\tau)}>=\frac{1}{3\tau}\int
|k||c(k)|^2 dk.
\end{equation}
From this equation we see, that this mean value is singular as it
is  in the classical theory.

The way to avoid singularity is to suggest, that the Universe was
burn not from a point but from a "seed" of "size" $a_0$. Then
expression for mean value changes to
\begin{eqnarray*}
<A(\tau)>=i\int \biggl(\frac{1}{a}\frac{\partial \psi}{\partial
a}(\hat A(\tau)\psi)^*~~~~~~~~~~~
\\
-\left(\frac{1}{a}\frac{\partial \psi}{\partial a }\right)^*\hat
A(\tau)\psi\biggr)d\phi\biggr|_{a=a_0}. \label{geizd}
\end{eqnarray*}
This lead to other question about underling theory giving size of
the seed.

Now we find solution for the Hamiltonian, containing the the
cosmological constant $V_0$:
\begin{equation}
\hat H=-\frac{1}{4}\left({\hat p_a}^2\frac{1}{a}+\frac{1}{a}{\hat
p_a}^2\right)+\frac{\hat p_\phi^2}{2 a^3}+a^3V_0.
\end{equation}
Evaluating commutators in a usual way we have
\begin{eqnarray}
(\hat a^3(\tau))^{\bm\cdot}=-\frac{3}{2}(\hat p_a  a+ a \hat p_a),\label{a1}\\
(\hat p_a  a+ a \hat p_a)^{\bm\cdot}=6\hat H_0-6Va^3,\label{a2}\\
6(\hat H_0-Va^3)^{\bm\cdot}=18V(\hat p_a  a+ a \hat
p_a).\label{a3}
\end{eqnarray}
 Form the equations (\ref{a2}) and (\ref{a3}) it follows that
$(\hat p_a  a+ a \hat p_a)^{\bm \cdot\bm\cdot}=18V_0(\hat p_a  a+
a \hat p_a)$ and $(\hat p_a  a+ a \hat p_a)=\hat D\,
\sinh(\sqrt{18V_0}\,\tau)+\hat B\, \cosh(\sqrt{18V_0}\,\tau)$,
where $\hat D$ and $\hat B$ are some operators not depending on
$\tau$.
 Finally arrive to
 \begin{widetext}
 \begin{equation}
\hat a^3(\tau)=a^3-\frac{3}{2}\left((\hat p_a  a+ a \hat
p_a)\frac{\sinh(\tau\sqrt{18V})}{3\sqrt{2V}}+\sqrt{\frac{2}{V}}
(\hat H_0-Va^3)(\cosh(\tau\sqrt{18V})-1)\right). \label{os}
 \end{equation}
 \end{widetext}

Similar solutions for Heisenberg operators, containing $\sinh$ and
$\cosh$ were obtained in \cite{lasuk,weist}.
 The operator given by (\ref{os}) is the local one. Thus
 to evaluate it mean value according to our rule (\ref{geiz}) we
 need to know the wave function only in the vicinity of
 $a=0$, so we can use the function (\ref{pack}), which is
 limiting value of the exact wave function at $a\rightarrow 0$.
Mean values take the form
\begin{eqnarray*}
<\hat a^3(\tau)>=\frac{\sinh (3\,\sqrt{2}\,{\sqrt{V}}\,\tau )}
  {\sqrt{2V}}\int \frac{\left( 3 + 2\,k^2 \right)
  }{2\,|k| }|c(k)|^2dk
\\
<\hat a^6(\tau)>=\frac{\left({\sinh
(3\,{\sqrt{2}}\,{\sqrt{V}}\,\tau )}\right)^2}{2\,V}\int k^2
|c(k)|^2dk.
\end{eqnarray*}
This shows, that dispersion $\frac{\sqrt{<\hat a^6>-<\hat
a^3>^2}}{<\hat a^3>}$ is not depends on $\tau$, exactly as in the
case of the free relativistic particle. Thus in this model
evolution of the Universe remains quantum all the time. This is
because we do not introduce here some fundamental length.  Such a
fundamental length appears if we take
$V(\phi)=\frac{m^2\phi^2}{2}$. One may suggest that expanding of
the Universe before the Compton wave length $1/m$ is quantum, and
when $<a(\tau)>$ becomes greater than $1/m$ the expansion can be
described classically. To see this explicitly we must to find
Heisenberg solutions with the above potential, but it is very
difficult problem.

\section{Operator ordering and the Universe rest mass}
In the previous consideration we use some particular operator
ordering in the equation (\ref{hamilt}). In the general case we
may write the Hamiltonian in the form
\begin{eqnarray}
\hat H_0=- \frac{1+M^2}{4}\left({\hat
p_a}^2\frac{1}{a}+\frac{1}{a}{\hat p_a}^2-\frac{2M^2}{1+M^2}\,\hat
p_a\frac{1}{a}\hat p_a\right)\nonumber\\
+\frac{\hat p_\phi^2}{2 a^3}.
\end{eqnarray}
To be consistent with the normalization procedure (\ref{nm})
parameter $M^2$ must be positive. The wave function satisfying to
the $H_0\psi=0$ is
\begin{equation}
\psi(a,\phi)=a^{1-i\sqrt{M^2+k^2}}e^{ik\phi}.
\end{equation}
We see that $M$ plays the role of the "rest mass" of the Universe.
The equations (\ref{a1}), (\ref{a2}), (\ref{a3}) holds also for
the $M\ne 0$ with the corresponding $H_0$. It allows us to
evaluate mean the values:
\begin{eqnarray*}
  <\hat a^3(\tau)>=\frac{ \sinh (3\,{\sqrt{2}}\,{\sqrt{V}}\,\tau )}
  {\,{\sqrt{ 2\,V}}}
  \int \frac{\left( 3 + 2\,k^2 + 2\,M^2 \right)}{2\sqrt{k^2 + M^2} }
  |c(k)|^2 d k,
\\
<\hat a^6(\tau)>=\frac{ \,\left({\sinh
(3\,{\sqrt{2}}\,{\sqrt{V}}\,\tau )}\right)^2}{2\,V}\int\left( k^2
+ M^2 \right)|c(k)|^2 d k.
\end{eqnarray*}

Taking into account that $\int |c(k)|^2 dk=1$ we can conclude that
if the Universe have rest mass much greater than the Plank mass
(equal to unity in our units) and the character value of the wave
vector $k$, then dispersion of the $a^3$ will be negligible,
exactly as in the case of the nonrelativistic particle, for which
dispersion of physical time is negligible.

\section{Conclusion}
We have considered quantum evolution of the Universe, originated
from the some fluctuation of the scalar field (wave packet).

\section*{ACKNOWLEDGMENT}
I am grateful to Dr. Vladimir Kalashnikov from Wein TU for deep
and valuable discussions.

\end{document}